%% file: defects5.tex
\newcommand{\be}{\begin{equation}}
\newcommand{\ee}{\end{equation}}
\newcommand{\ba}{\begin{eqnarray}}
\newcommand{\ea}{\end{eqnarray}}
\newcommand{\bi}{\begin{itemize}}
\newcommand{\ei}{\end{itemize}}

\documentclass[11pt]{JHEP3}
\usepackage{graphicx}
\usepackage{amssymb}
\usepackage{epsfig}
\input epsf.tex

\include{newcomm}

\title{Looking for defects in the 2PI correlator}

\author{Arttu Rajantie$^{a}$ and Anders Tranberg$^{b,c}$\\
$^{a}$ Theoretical Physics, Blackett Laboratory,
 Imperial College,  London SW7 2AZ, United Kingdom.\\
$^{b}$ Department of Physics \& Astronomy, University of
Sussex, Falmer, Brighton, BN1 9QH, United Kingdom.\\
$^{c}$ DAMTP, University of Cambridge, Wilberforce Road, Cambridge CB3 0WA, United Kingdom.
} 


\preprint{Imperial/TP/06/AR/1, DAMTP-2006-57,}

\abstract{
Truncations of the 2PI effective action are seen as a promising way of studying non-equilibrium dynamics in quantum field theories. We probe their applicability in the non-perturbative setting of topological defect formation in a symmetry-breaking phase transition, by comparing full classical lattice field simulations and the 2PI formulation for classical fields in an O($N$) symmetric scalar field theory. At next-to-leading order in $1/N$, the 2PI formalism fails to reproduce any signals of defects in the two-point function.
This
suggests that one should be careful when applying the 2PI formalism
for symmetry breaking phase transitions.}

\begin{document}
%\maketitle

%\newpage
  
%%%%%%%%%%%%%%%%%%%%%%%%%%%%%%%%%%%%%%%%%%%%%%%%%%%%%%%%%%%%%%%%%%%%%

% SECTION INTRODUCTION

\section{Introduction\label{introduction}}
An exact numerical solution of non-equilibrium dynamics in quantum field theories is only possible in the very simplest cases, in contrast to vacuum or thermal equilibrium. 
One generally has to resort to drastic approximations.
In the classical
approximation~\cite{Khlebnikov:1996mc,Prokopec:1996rr}, the Hamiltonian equations of motion are solved for an
ensemble of initial conditions, and observables are averaged over the
ensemble. The approximation includes non-linear and non-perturbative
effects, but its applicability is restricted to cases when
field occupation numbers are large, such as high temperatures.

An alternative approach is to study the evolution of the correlators
themselves using Schwinger-Dyson equations in real time. These require
truncation of an infinite series of diagrams, ordered
according to some expansion of choice. One organisation of these
diagrams follows elegantly from variations of the 2PI (or $n$PI)
effective action, from which results Schwinger-Dyson equations for the
1- and 2- (and up to $n$-) point correlators \cite{Cornwall:1974vz}.

Unlike naive perturbation theory,
truncations of the Schwinger-Dyson equation based on the 2PI
effective action conserve energy in time, 
making the approach suitable for studying the dynamics over
extended time intervals (see for instance \cite{Berges:2001fi}). 
It has, for instance, been used to investigate equilibration and
thermalisation of both scalar and fermion fields
\cite{BeCo01,Aarts:2001qa,Berges:2002wr,Juchem:2003bi,Arrizabalaga:2005tf}, reheating after inflation \cite{Berges:2002cz,Arrizabalaga:2004iw} and to
calculate transport coefficients
\cite{Aarts:2003bk,Aarts:2004sd,Aarts:2005vc} and critical exponents \cite{Alford:2004jj}. It is renormalisable \cite{VanHees:2001pf,vanHees:2002bv,Blaizot:2003br,Blaizot:2003an,Berges:2005hc} for all truncations, and is in particular applicable for very late times,
where the classical approximation will break down as systems
equilibrate classically rather than quantum \cite{Arrizabalaga:2004iw}.

In this paper, we will concentrate on the formation of topological defects (or solitons \cite{Manton:2004tk}) in
a symmetry breaking phase transition~\cite{Kibble:1976sj,Rajantie:2001ps}. This is a highly non-perturbative
non-equilibrium process.
It also has physical relevance,
because topological defects are produced in phase transitions of various
condensed matter systems (see, for instance \cite{Rajantie:2001ps} and references therein) and they may have also been formed in the early universe, at the
end of inflation~\cite{Felder:2000hj,Copeland:2002ku}. While it would be interesting to study this process in
quantum field theory, we restrict ourselves to a classical field theory.
By using the classical limit of the 2PI equations~\cite{Aarts:2001yn},
we can compare the results of the 2PI approach to the full numerical solution of the classical equations of motion, which is in principle exact.

In a classical simulation, topological defects can be easily detected in
the resulting field configurations, but in the 2PI approach, the
basic quantity 
available to us is the fully averaged two-point function. 
Therefore, we first show how the presence of defects, kinks for $N=1$ and textures for $N=2$, manifests itself in the two-point function. 

In section \ref{modelanddefects} we introduce the model we are considering, an O($N$) model of scalar fields in 1+1
dimensions. In section \ref{lookingfordefects} we derive tell-tale
signals in the equal-time two-point correlator of the presence of defects.
Then in section \ref{fullclassical} we perform detailed simulations
of the full classical theory and establish how these signals manifest
themselves. The 2PI formalism is then intoduced in section \ref{2piformalism} and
 we compare numerical simulations of the LO and NLO approximations to the full
classical result. We conclude in section \ref{outlook}.

% SECTION MODEL AND DEFECTS

\section{Model and defects\label{modelanddefects}}
We consider a classical theory of $N$ real scalar fields $\phi_{a}$ with $a\in\{0,..,N-1\}$  in $1+1$
dimensions. The continuum action is
\be
S=\int dx\,dt\,\left(\frac{1}{2}\partial_{\mu}\phi_{a}\partial^{\mu}\phi_{a}-
V(\phi_a)
\right),
\ee
and has an O($N$) symmetry.
We shall investigate the dynamics of the system in a simple setup in which the potential varies with time. Initially, it corresponds to
a free field with mass  $\mu$, 
\be
\label{equ:Vini}
V_{\rm ini}(\phi_a)=\frac{1}{2}\mu^2\phi_a\phi_a,
\ee
and at time $t=0$, it changes instantaneously to
\be
\label{equ:Vbroken}
V(\phi_a)=
-\frac{1}{2}\mu^{2}\phi_{a}\phi_{a}+\frac{\lambda}{24N}(\phi_{a}\phi_{a})^{2},
\ee
triggering a symmetry breaking transition. 
We add a small constant damping term $\Gamma\partial_{t}\phi$ to the equation of motion,
so that it becomes
\be
\label{equ:eom}
\partial_{t}^{2}\phi_{a}(x,t)+\Gamma\,\partial_{t}\phi_{a}(x,t)-\partial_{x}^{2}\phi_{a}(x,t)-\mu^{2}\phi_{a}(x,t)+\frac{\lambda}{6N}\left(\phi_{b}\phi_{b}\right)\phi_{a}(x,t)=0.
\ee
This ensures that the system reaches eventually a zero-temperature state with the O($N$) symmetry broken spontaneously and 
$\phi_a\phi_a=(6N/\lambda)\mu^2\equiv v^2$.
Without damping the system would equilibrate in a state with a non-zero temperature, and the symmetry would be restored.
The equation of motion (\ref{equ:eom}) is discretised in the spatial direction on a lattice of spacing $a$ and in the temporal direction using the leapfrog algorithm with time step $\delta t$ for the time derivative.

For $N=1$, the model has two degenerate vacua at $\phi_{0}=\pm
v=\pm\sqrt{6/\lambda}\mu$, and there are
topological defects, kinks, which interpolate
between them. The classical kink solution is 
\be
\label{equ:kinkprofile}
\phi_{\rm kink}(x)=v\tanh\frac{x}{d},
\ee
where $d=\sqrt{2}/\mu$ is the kink thickness.

For $N=2$, the vacuum manifold is a circle, and the model does not have localised defect solutions. 
However, there are textures %(skyrmions) 
which correspond to field configurations 
that wind around the vacuum manifold. In infinite volume, they are unstable
against growing to infinite size and becoming indistinguishable from the vacuum. 
However, if the system has a finite size $L$, the classical theory has stable texture solutions~\cite{Davis:1986nr}
\be
\label{equ:texture}
\Phi(x)\equiv
\phi_0(x)+i\phi_1(x)
=\Phi_{\rm text}^{N_w}(x)\equiv ve^{2\pi i N_wx/L},
\ee
where the winding number $N_w$ is an integer.
Because textures are stabilised by a finite energy barrier rather than a fundamental conservation law, they are in fact 
only metastable in the presence of 
thermal or quantum fluctuations.
For $N>2$ there are no topological defects. 

We choose the initial conditions at time $t=0$ to mimic the quantum vacuum state corresponding to the potential (\ref{equ:Vini}). Because this is a free theory, the equal-time quantum
two-point functions of the field $\phi_{a}$ and its canonical momentum 
$\pi_{a}=\partial_{t}\phi_{a}$ are simply
\ba
\label{equ:initcond}
\langle
\phi_{a}(k)\phi_{b}(q)\rangle&=&(2\pi)\delta(k+q)
\delta_{ab}\frac{1}{2\omega_k},\nonumber\\
\langle
\pi_{a}(k)\pi_{b}(q)\rangle&=&(2\pi)\delta(k+q)
\delta_{ab}\frac{\omega_k}{2},\nonumber\\
\langle
\phi_{a}(k)\pi_{b}(q)\rangle&=&0,
\ea
where  $\omega_k=\sqrt{k^2+\mu^2}$.
Our initial conditions are given by a Gaussian ensemble of field configurations
which has these same two-point functions.
This choice of initialisation has
been used extensively in the study of inflationary reheating in
cosmology~\cite{Khlebnikov:1996mc,Prokopec:1996rr}, also using the 2PI
formalism \cite{Arrizabalaga:2004iw}
In our case, it is not important how well these initial conditions reproduce the actual quantum dynamics, since we are only interested in the classical dynamics. For the 2PI formalism, the initial ensemble has to be Gaussian and this is a simple and convenient choice.

The classical equation of motion (\ref{equ:eom}) 
allows us to rescale the coupling $\lambda$ to unity, suggesting that only the dimensionless ratio $\lambda/\mu^2$ plays a role. However,
the initial conditions (\ref{equ:initcond}) remove this freedom.
We therefore keep both
the coupling and the mass parameter. %They will, however not be varied.

We solve the time evolution of the system using the initial conditions
(\ref{equ:initcond}) using two different approaches: the full
classical equations, and the 2PI formalism. In the former case, the
classical equation of motion (\ref{equ:eom}) is solved numerically for
a large number of initial configurations that are drawn from the
distribution specified by Eq.~(\ref{equ:initcond}). Apart from the
discretisation and finite-size errors, which should be  similar in both cases, the only error in the full classical approach is statistical, due to a finite number of initial conditions. In contrast, there is no statistical error in the 2PI approach, but the truncation of the Schwinger-Dyson equation introduces a systematic
error.

% SECTION LOOKING FOR DEFECTS IN THE PROPAGATOR

\section{Looking for defects in the propagator\label{lookingfordefects}}

% SUBSECTION KINKS

\subsection{Kinks\label{kinks}}
In classical simulations, it is easy to study topological defects by considering
individual realisations of the field. Each realisation is inhomogeneous, and a kink, for instance, corresponds to a point where $\phi_0$ vanishes.
In contrast, the 2PI formalism does not give information about individual realisations, and if the initial ensemble is symmetric, the mean field remains
zero $\langle\phi_a\rangle=0$. All information about the system is encoded in
the full two-point correlator
$G_{ab}(x,y,t,t')=\langle\phi_{a}(x,t)\phi_{b}(y,t')\rangle$,
which is invariant under O($N$) transformation and spatial translations, $G_{ab}(x,y,t,t')=\delta_{ab}G(|x-y|,t,t')$,
and corresponds to an average over the whole ensemble.
Therefore, we need to know what effect topological defects have on the
correlator.
Once we know that, we can calculate this quantity in full classical
simulations, averaging over an ensemble of realisations, and in the 2PI formalism using different truncations.

A popular approach \cite{Liu:1992ey,Ibaceta:1998yy,Antunes:2005bx} is to assume that the field 
$\phi_a$ is Gaussian. In that case, the density of zeros of the field is
given by
\be
\label{zeroscorr}
n_0=\frac{1}{\pi}\sqrt{-\frac{G''(0,t,t)}{G(0,t,t)}}.
\ee
Since the zeros can be identified with kinks, this would then give the number density of kinks, $n=n_0$.
However, since $\phi_a$ becomes non-Gaussian soon after the
transition, this  approach does not work~\cite{Ibaceta:1998yy}. Instead, we need a method that works for strongly non-Gaussian
fields.

Another approach advocated in \cite{Stephens:1998sm} is to 
look for a characteristic length scale $L_{\rm max}$ in the propagator
and identify it with the typical distance between kinks, so that
$n\approx 1/L_{\rm max}$. It is presumably very generally true that a
given number density $n$ of kinks does indeed introduce  some feature at the corresponding length scale. However,
the existence of a characteristic length scale does not imply that there
are kinks in the system, and this approach is therefore not suitable for our purposes.

Let us, instead, calculate directly what the form of the two-point function is in the presence of defects. We consider a one-dimensional lattice with spacing $a$ and assume that there are randomly distributed infinitesimally 
thin kinks with number density $n$. Let us further assume that there are no other fluctuations, so that the field only takes values $\pm
v$. The changes of sign correspond to the locations of the kinks. 
We can choose the field to be positive at point $x$. 

Clearly $G(0,t,t)=v^2$. 
The probability of there being a kink between points $x$ and $x+a$ is $na$, and
therefore the neighbouring point has the value $\phi_0(x+a)=-v$ with
probability $na$, and $\phi_0(x+a)=+v$ with probability $1-na$. As a result $G(a,t,t)=(1-2na)v^2$.
At distance $2a$ we have $\phi(x+2a)=+v$ if there are either two or zero kinks between the points, whereby
$G(2a,t,t)=\left((1-na)^{2}+na^{2}-2na(1-na)\right)v^2$. In general,
\ba
\label{kinkworkout}
G(Ma,t,t)&=&v^2\sum_{k=0,..,M}(-1)^{k}(na)^{k}(1-na)^{M-k}\frac{M!}{(M-k)!k!}=v^2(1-2na)^M\nonumber\\
&\stackrel{a\rightarrow0}{\longrightarrow}&v^2e^{-2n|x-y|},
\ea
where we have kept $|x-y|=Ma$ constant when taking the limit $a\rightarrow 0$.
The Fourier transform of the correlator is
\be
\label{equ:porod}
G(k,t,t)= \frac{4n}{4n^2+k^{2}}v^2.
\ee
We therefore expect that the correlator has this form in the presence of kinks. 
A fit of this form allows us to determine the kink density $n$. 
Note, however, that since we assumed that the kinks are infinitesimally thin, this expression is only valid at distances longer than the kinks thickness, i.e., $k\ll \mu$. 

Note also that the correlator has a power-law form $G(k,t,t)\propto k^{-2}$ at intermediate length scales $n \ll k \ll \mu$.
This is a special case of what is known as a ``Porod tail'' in the
correlator. For general $N$ and spatial dimension $D$, $G({\bf
  k},t,t)\propto k^{-(N+D)}$ (see for instance \cite{Bray}). 
  
Unfortunately, Eq.~(\ref{equ:porod}) by itself is not an ideal indicator of the presence of defects,
because the correlator of a weakly-coupled scalar field in thermal equilibrium has the same
form,
\be
\label{thermsign}
G_{\rm therm}(k,t,t)=\frac{T}{k^2+m^2}.
\ee
Therefore, we need to take into account the finite thickness of the kinks.
This gives an extra multiplicative factor to the correlator,
\be
\label{kinksig}
G(k,t,t)=\frac{4n}{4n^2+k^{2}}\frac{k^2}{4}|\phi_{\rm kink}(k)|^2,
\ee
where $\phi_{\rm kink}(k)$ is the Fourier transform of the kink profile.
In the absence of any fluctuations, this expression should be valid
for $k\ll 1/a$. 

For the kink solution in Eq.~(\ref{equ:kinkprofile}), we obtain
\be
\label{kinksolution}
\phi_{\rm kink}(k)=
{{2iv}\over{k}}\frac{\frac{1}{2}\pi kd}{\sinh \frac{1}{2}\pi kd},
\ee
and therefore we should find
\be
\label{equ:kinksign}
G(k,t,t)=\frac{4n}{4n^2+k^{2}}\left(
\frac{\frac{1}{2}\pi kd}{\sinh \frac{1}{2}\pi kd}\right)^2
v^2.
\ee
Since $d=\sqrt{2}/\mu$ and $v=\sqrt{6/\lambda}\mu$ are known, this expression
has only one free parameter $n$. In the presence of thermal or quantum fluctuations, these parameters and the kink profile would change. In fact, one could use Eq.~(\ref{kinksig}) to measure the kink shape $\phi_{\rm kink}(k)$.

Using Eq.~(\ref{equ:kinksign}) we can also calculate, which value the Gaussian formula Eq.~(\ref{zeroscorr}) would give for the kink density. A straightforward calculation shows that in the limit $nd\ll 1$,
\be
n_0=\frac{1}{\pi}\sqrt{
\frac{\int dk k^2 G(k)}{\int dkG(k)}}
\approx\frac{2}{\pi}\sqrt{\frac{n}{3d}}.
\ee
Since this result is not even proportional to $n$ but is sensitive to the kink shape, it is clear that Eq.~(\ref{zeroscorr}) cannot be used to measure kink density.

% SUBSECTION TEXTURES
   
\subsection{Textures\label{textures}}
As is obvious from Eq.~(\ref{equ:texture}), textures are pure plane wave
configurations, and therefore their signature in the two-point correlator is
very simple. In a finite system with length $L$, a texture with winding $N_w$ gives a contribution $Lv^2/2$ to the correlator at wave number $k=2\pi N_w/L$.
Thus, if textures with winding $N_w$ appear with probability $p(N_w)$ in
the ensemble, the correlator will simply be
\be
\label{texturesolution}
G(k,t,t)=p\left(\frac{kL}{2\pi}\right)\frac{Lv^2}{2}.
\ee

We can even derive the shape of the probability distribution $p(N_w)$, if we
assume that the textures are formed by the Kibble mechanism~\cite{Kibble:1976sj}.
Immediately after the transition, the field $\phi_a$ is more or less constant at distances less than some ``freeze-out'' length scale $\xi$, and completely uncorrelated at longer distances. In length $L$, there are therefore $N_\xi=L/\xi$ uncorrelated 
regions, which we label by $i=1,\ldots,N_\xi$. In each of them, the field has some constant value,
\be
\label{N2field}
\Phi_i=v e^{i\theta_i},
\ee
where the phase angle $\theta_i$ is random. The field interpolates smoothly
between these values when we go from one region to the next, and it is
natural to assume that it follows the shortest path on the vacuum manifold. 
The phase angle changes by an amount $\Delta\theta_i=[\theta_{i+1}-\theta_i]_\pi$,
where the subscript $\pi$ indicates that we choose it to be in the range $-\pi<\Delta\theta_i\le \pi$.
When we follow the field throughout the whole system, through the boundary, back to the original point, the phase angle may wind around the vacuum manifold some
number of times $N_w$ given by
\be
\label{phases}
\sum_{i=1}^{N_\xi}\Delta\theta_i=2\pi N_w.
\ee 
When we let the system evolve after the transition, 
the winding number $N_w$ remains unchanged and the system equilibrates in the 
local minimum, which is the texture solution with winding $N_w$.

To estimate $p(N_w)$, we note that the changes $\Delta\theta_i$ of the phase angle
are random with a uniform probability distribution in $(-\pi,\pi]$.
According to the central limit theorem, the probability distribution 
in the limit $N_\xi\rightarrow\infty$ is Gaussian,
\be
\label{textGauss}
p(N_w)=\sqrt{\frac{6}{\pi N_\xi}}
\exp\left(-\frac{6N_w^2}{N_\xi}\right).
\ee
The correlator should therefore be
\be
\label{equ:textsign}
G(k,t,t)=v^2\sqrt{\frac{3\xi L}{2\pi}}\exp\left(-\frac{3\xi L}{2\pi^2}
k^2\right).
\ee
The only free parameter in this expression is $\xi$. 

Again, we have to be careful when comparing this result with measurements, 
because one might expect the correlator to be Gaussian at very low $k$
in any case. This is because massless Goldstone modes with $k<\Gamma/2$ are
overdamped and will decay as $\exp(-(k^2/\Gamma)t)$, giving
\be
\label{propGoldstone}
G(k,t,t)\sim\exp\left(-\frac{t}{\Gamma}k^2\right).
\ee
A Gaussian shape by itself would therefore not be evidence for the presence 
of textures, and therefore we have to check that the exponent approaches a constant at $t\rightarrow\infty$ rather than growing linearly as Eq.~(\ref{propGoldstone}) would predict.

% SECTION FULL CLASSICAL SIMULATIONS

\section{Full classical simulations\label{fullclassical}}

We discretised the equations on a lattice with spacing $a=1$, which means that
all dimensionful quantities are expressed in lattice units. The time step was 
 $\delta t=0.1$. In all runs, we used the coupling $\lambda=0.6$.

% SUBSECTION KINKS

\subsection{Kinks\label{simkinks}}

\FIGURE{
\epsfig{file=pictures/smoothkink.eps,width=12cm,clip}
\caption{
The two-point function $G(k,t,t)$ for $N=1$ at time $t=10000$. The dashed curve
is a fit of the form (\ref{equ:kinksign}) with $n=116(1)\times 10^{-5}$.
The inset shows the number density of kinks measured in three different ways: Counting zeros of $\phi$ directly from field configurations (solid line), fitting Eq.~(\ref{equ:kinksign}) to the measured correlation function (dashed line) and using the Gaussian formula in Eq.~(\ref{zeroscorr}) (dotted line).
While Eq.~(\ref{zeroscorr}) works well at early times when the field configuration is Gaussian, it fails at later times when the kinks have actually formed. In contrast,  Eq.~(\ref{equ:kinksign}) agrees very well with the direct measurement at late times.
}
\label{fig:smoothkink}
}
To test the prediction (\ref{equ:kinksign}) we carried out 2000 runs 
with $\mu^2=0.01$ on a
lattice with $L=16384$. 
The damping rate was $\Gamma=0.001$.
The kink thickness in this case is $d\approx 14.14\gg 1$,
and the kinks should therefore be well approximated by the continuum solution
(\ref{equ:kinkprofile}). 

In Fig.~\ref{fig:smoothkink} we show the two-point function measured at time $t=10000$, together with a fit of the form (\ref{equ:kinksign}). The only free parameter in the fit is the kink number density $n=0.00116(1)$. Only points at $k<0.15$ were included in the fit. At higher $k$, the correlator is still dominated by the perturbative ``quantum'' fluctuations, 
which decay exponentially with time because of the damping term. 

The fit is very good at $k<0.15$ (i.e., $k/\mu<1.5$), which shows that the correlator is
dominated by kinks of the form (\ref{equ:kinkprofile}).
Further evidence for this is shown in the inset of Fig.~\ref{fig:smoothkink}, where we compare
the number density of kinks determined by fitting the measured two-point function with Eq.~(\ref{equ:kinksign}) to the result obtained by counting the
zeros of the field in the lattice field configurations. The two results agree
perfectly. For comparison, we also show the incorrect result obtained with Eq.~(\ref{zeroscorr}).

% SUBSECTION TEXTURES

\subsection{Textures\label{simtextures}}
\FIGURE{
\epsfig{file=pictures/textures.eps,width=12cm,clip}
\caption{The two-point function $G(k,t,t)$ for $N=2$ at times $t=1000$ (white), $t=3000$ (grey) and $t=10000$ (black).
The curves are Gaussian fits of the form (\ref{equ:textsign})
with $\xi=7.23\pm0.02$, $\xi=13.03\pm0.04$ and $\xi=19.94\pm0.22$.
In the inset, the data points show the time evolution of the fit parameter $\xi$, and
the curve is an exponential fit.
}
\label{fig:textures}
}

To test Eq.~(\ref{equ:textsign}), we chose $L=4096$, $\mu^2=0.49$ and $\Gamma=0.01$.
Fig.~\ref{fig:textures} shows the correlator at various times, together
with a fit of the form (\ref{equ:textsign}). 
Again, the fit is very good.
The inset shows the fit parameter $\xi$ 
at various times, together with an exponential fit
\be
\label{textexp}
\xi(t)=\xi(\infty)-\Delta\xi e^{-\alpha t}.
\ee
The fit parameters are $\xi(\infty)=21.4(2)$, $\Delta\xi=18.4(1)$ and
$\alpha=2.68(6)\times 10^{-4}$.
This shows that the length scale $\xi$ approach asymptotically a constant
at late times, ruling out the diffusive behaviour of Eq.~(\ref{propGoldstone}) and providing a clear signal for the presence of textures.

% SUBSECTION HIGHER H

\subsection{Higher $N$\label{higherN}}
\FIGURE{
\epsfig{file=pictures/N4h.eps,width=12cm,clip}
\caption{The two-point function $G(k,t,t)$ for $N=4$ at times $t=1000$ (white), $t=3000$ (grey) and $t=10000$ (black).
The curves are exponential fits to the first six data points. The plateau 
(or $k^{-1}$ behaviour to be more precise) at
higher $k$ is a remnant of the initial ``quantum'' fluctuations (\ref{equ:initcond}).
In the inset, the data points show the time evolution of the exponent, and
the blue curve is a power-law fit with exponent $0.670(6)$.
}
\label{fig:N4}
}

For $N>2$, there are no topological defects, and the system should end up in
the vacuum state, in which the two-point function is simply a delta function.
In finite volume,
\be
\label{highNlimit}
\lim_{t\rightarrow\infty}G(k,t,t)=\left\{\begin{array}{ll} v^2L/N & \mbox{for}~k=0,\cr
0 & \mbox{fot}~k\ne 0.\end{array}\right.
\ee
In Fig.~\ref{fig:N4}, we show the correlator for $N=4$. 
It appears to be well fitted by an exponential, 
\be
\label{highNfit}
G(k,t,t)\propto e^{-lk}.
\ee
As the
inset shows, the exponent grows as $l\propto t^{2/3}$,
becoming singular at $t\rightarrow\infty$ as expected.

% SECTION 2PI-1/N APPROXIMATION at NLO

\section{2PI-1/N approximation at LO and NLO\label{2piformalism}}

The 2PI formalism is based on quantum field theory, but it can be used to
study an ensemble of classical field configurations. The corresponding 
equations are obtained by taking the classical limit $\hbar\rightarrow 0$
of the quantum equations. In our case, this is not entirely straightforward because our classical
equation of motion (\ref{equ:eom}) contains a damping term, and it is therefore not immediately obvious what the corresponding quantum theory should be.

To avoid this problem, we note that Eq.~(\ref{equ:eom}) is also the equation
of motion of an undamped 2+1-dimensional scalar field theory in an expanding anisotropic
space with metric 
\be
\label{metric}
ds^{2} = dt^{2}-dx^{2}-a(t)^{2}dy^{2}.
\ee
The equation of motion for the scalar fields $\phi_a$ is
\be
\label{eom2d}
\partial_{t}^{2}\phi_{a}+\frac{\dot a}{a}\partial_{t}\phi_{a}-\partial_{x}^{2}\phi_{a}(x,t)-\frac{1}{a^2}\partial_{y}^{2}\phi_{a}(x,t)+V'(\phi_a)=0.
\ee
If the scale factor grows exponentially,
\be
\label{scalefactor}
a(t) = a(0)\,e^{\Gamma t},
\ee
and all the fields $\phi_a$ are independent of $y$, this reduces to Eq.~(\ref{equ:eom}). However, since the dynamics is undamped, the quantum generalisation of the system is obvious. 

To derive the 2PI equations, we write the action as\footnote{The action
is in principle defined on the Keldysh contour $\tilde{S}=\tilde{S}_\mathcal{C}$. We will
suppress this complication for the moment, since it has no
implications when deriving the classical equations of motion. It does
enter and is crucial in the derivation of the 2PI evolution equations (see also Appendix \ref{app2PI}).}
\be
\label{action}
\tilde{S}=\int a(t)dy\,dx\,dt\,\left(\frac{1}{2}\partial_{t}\phi_{a}\partial_{t}\phi_{a}-\frac{1}{2}\partial_{x}\phi_{a}\partial_{x}\phi_{a}-\frac{1}{2\,a^{2}(t)}\partial_{y}\phi_{a}\partial_{y}\phi_{a}-V(\phi_{a})\right).
\ee
Imposing that the field is explicitly independent
of the $y$-coordinate $\phi(x,y,t)=\phi(x,t)$, we have
\be
\label{action2}
\tilde{S}=L_{y}\int a(t)dx\,dt\,\left(\frac{1}{2}\partial_{t}\phi_{a}\partial_{t}\phi_{a}-\frac{1}{2}\partial_{x}\phi_{a}\partial_{x}\phi_{a}-V(\phi_{a})\right),
\ee
where we have performed the integration over $y$, $\int dy = L_{y}$. We
end up with an effective theory of $N$ real scalar fields in $1+1$ dimensions,
but with a non-standard action, including a time-dependent
factor $a(t)$. In the classical limit, the factor $L_y$ drops out, so we can ignore it.

We note that we can write the
action as 
\be
\tilde{S}=\int dx\,dt\,\left(-a(t)\phi_{a}\left[\partial_{t}^{2}+\Gamma\,\partial_{t}-\partial_{x}^{2}-\mu^{2}\right]\phi_{a}-\frac{a(t)\lambda}{24N}(\phi_{a}\phi_{a})^{2}\right).
\ee

In the 2PI formalism,\footnote{For details of the implementation of the 1/N expansion applied to the O($N$)
model see \cite{Berges:2001fi,Aarts:2002dj} and Appendix \ref{app2PI}.} evolution equations for the correlator
$G(x,y,t,t')$ are derived by varying the 2-point irreducible (2PI)
effective action corresponding to Eq.~(\ref{action2}). The result is the Schwinger-Dyson equation
for the statistical (F) and spectral ($\rho$) parts of the propagator
$G(x,y,t,t')=F(x,y,t,t')-\frac{i}{2}{\rm sign}(t,t')\rho(x,y,t,t')$. We assume
that the mean field is zero and explicitly impose
O($N$)-symmetry $G_{ab}=\delta_{ab}G$, as well as homogeneity and
isotropy, $G(x,y,t,t')=G(r,t,t')$, $r=|x-y|$, so that
\ba
\label{schwingerdyson}
\left(\partial_{t}^{2}-\partial_{x}^{2}+\Gamma\partial_{t}+M^{2}(t)\right)F(r,t,t')&=&-\int_{0}^{t}dz\,dt''\,\Sigma_{\rho}(r-z,t,t'')a(t'')F(z,t'',t')\nonumber\\&&+\int_{0}^{t'}dz\,dt''\,\Sigma_{F}(r-z,t,t'')a(t'')\rho(z,t'',t'),\\
\left(\partial_{t}^{2}-\partial_{x}^{2}+\Gamma\partial_{t}+M^{2}(t)\right)\rho(r,t,t')&=&-\int_{t'}^{t}dz\,dt''\,\Sigma_{\rho}(r-z,t,t'')a(t'')\rho(z,t'',t'),
\ea
\ba
\label{auxeom}
\frac{3N}{\lambda}D_{F}(r,t,t')&=-\Pi_{F}(r,t,t')&+\int_{0}^{t}\,dz\,dt''\,\Pi_{\rho}(r-z,t,t'')a(t'')D_{F}(z,t'',t')\nonumber\\&&-\int_{0}^{t'}\,dz\,dt''\,\Pi_{F}(r-z,t,t'')a(t'')D_{\rho}(z,t'',t'),\\
\frac{3N}{\lambda}D_{\rho}(r,t,t')&=-\Pi_{\rho}(r,t,t')&+\int_{0}^{t}\,dz\,dt''\,\Pi_{\rho}(r-z,t,t'')a(t'')D_{\rho}(z,t'',t'),
\ea
with the effective mass given by the local part of the self-energy
\be
\label{effmass}
M^{2}(t)=-\mu^{2}+\lambda\frac{N+2}{6N}F(0,t,t).
\ee
The objects $D_{F}$ and $D_{\rho}$ are components of an auxiliary
propagator \cite{Aarts:2002dj}. The non-local parts of the self-energies 
are given at NLO of the 1/N expansion as 
\ba
\Sigma_{F}(r,t,t')&=&-\frac{\lambda}{3N}\left[F(r,t,t')D_{F}(r,t,t')\right],\\
\Sigma_{\rho}(r,t,t')&=&-\frac{\lambda}{3N}\left[\rho(r,t,t')D_{F}(r,t,t')+F(r,t,t')D_{\rho}(r,t,t')\right],\\
\Pi_{F}(r,t,t')&=&-\frac{N}{2}\left[F(r,t,t')F(r,t,t')\right],\\
\Pi_{\rho}(r,t,t')&=&-NF(r,t,t')\rho(r,t,t').
\label{sigmaNLO}
\ea
We have taken the classical limit, as prescribed in \cite{Aarts:2001yn}.

The functions
$F$($\rho$), $\Sigma_{F}$($\Sigma_{\rho}$), $\Pi_{F}$($\Pi_{\rho}$)
and $D_{F}$($D_{\rho}$)
are (anti-)symmetric in $t,t'$ and in particular
$\rho(t,t')|_{t=t'}=0$. When $\Gamma=0$ we should set
\be
\label{rhonodamp}
\partial_{t}\rho(t,t')|_{t=t'}=1,
\ee
for all times. With the damping term, we find instead
\be
\label{rhodamp}
\partial_{t}\rho(t,t')|_{t=t'}\propto e^{-\Gamma t}.
\ee
The scale factor $a(t)$ only appears in the self-energy, for every
time slice $t''$ taking on the appropriate value $a(t'')$. 

At leading order (LO), $N\rightarrow\infty$, we should drop all the
non-local self-energies (right-hand sides of
Eq.~(\ref{schwingerdyson})) and take the limit 
\be
\label{mLOlimit}
\frac{N+2}{6N}\rightarrow \frac{1}{6},
\ee
in Eq.~(\ref{effmass}). Keeping all of (\ref{effmass}) constitutes the
Hartree aproximation, which is leading order in a coupling expansion,
but a mixture of LO and some NLO in the $1/N$ expansion. Clearly,
since at LO the equations have no reference to $N$, it is impossible for
the approximation to know about defects. The Hartree approximation
amounts to a rescaling of the coupling $\lambda\rightarrow \lambda
(N+2)/N$, and so is equivalent to LO \footnote{Note, that in the
presence of a mean field $\langle\phi\rangle=v$, at LO the
fluctuations around this mean field satisfy Goldstone's theorem (and
so has zero modes for $N>1$), whereas the Hartree approximation does not. When
$\langle\phi\rangle=0$, Hartree acts as LO with the rescaled coupling,
and hence finds a different value for $v^{2}\rightarrow \frac{N}{N+2}v^{2}$ ($F(r=0)=v^{2}N/(N+2)$)
around which it again has zero modes.}.

\subsection{Spinodal transition in the free and LO approximations\label{LO}}

At very early times, we can neglect the coupling altogether, and solve
for the correlator as the spinodal transition proceeds,
\ba
\label{freepropmain}
\langle\phi_{\bf k}(t)\phi_{\bf k}^{\dagger}(t')\rangle=
\frac{e^{-\Gamma (t+t')/2}}{4\,\omega_{k}^{+}}&&
\left[
\left(1+\frac{\tilde{\omega}_{k}^{2}}{\omega_{k}^{2}}\right)\cos[\omega_{k}(t-t')]+
\left(1-\frac{\tilde{\omega}_{k}^{2}}{\omega_{k}^{2}}\right)\cos[\omega_{k}(t+t')]
\right.\nonumber\\
&&\left.
-2i\left(\frac{\omega^{+}_{k}}{\omega_{k}}\sin[\omega_{k}(t-t')]+i\tilde{\Gamma}\sin[\omega_{k}(t+t')]\right)
\right].
\ea
where (see Appendix A), $\omega_{k}^{\pm}=\sqrt{k^{2}\pm\mu^{2}}$, and
\ba
\omega_{k}^{2}=k^{2}-\left(\mu^{2}+\Gamma^{2}/4\right),\qquad
\tilde{\omega}_{k}^{2}=k^{2}+\left(\mu^{2}+\Gamma^{2}/4\right),\qquad \tilde{\Gamma}=\frac{\Gamma}{2\,\omega_{k}}.
\ea
Fig.~\ref{freeandLOF} (left) shows the evolution of the equal time,
equal space correlator at early
times in the free, LO and NLO approximations, as well as for the full
classical simulation. The initial growth is
faster than
exponential. The back-reaction of the interaction term
ends the growth at (N)LO, and the correlator starts oscillating around
it's vev. The apparent damping at LO is the effect of different
frequency modes coming out of phase (``dephasing''). No memory is lost
through dissipation, and (partial) recurrence is seen at later times, when the
(most dominant) modes come back into phase. At NLO the damping is
real, and the system will eventually thermalise \cite{Arrizabalaga:2004iw}.

\FIGURE{
\epsfig{file=./pictures/Ftt.eps,width=7cm,clip}
\epsfig{file=./pictures/freeLoc.eps,width=7cm,clip}
\caption{Left: Equal time correlator $F(t,t,r=0)$ in time for the
free, LO, NLO approximations and the full classical simulation. Right: The correlator $\ln(G(k,t,t))$ vs  $k$ in
the free (solid lines) and LO (dashed lines) approximations. $n_x=256$, $\lambda=0.6$, $\Gamma=0.02$, $\mu^{2}=0.49$.}
\label{freeandLOF}
}

Fig.~\ref{freeandLOF} (right) shows the equal time correlator in momentum
space as the tachyonic transition takes place. Around time $t=4$, the
back reaction kicks in and growth ends. There is a clear separation
between unstable $k<\mu$ and oscillating modes $k>\mu$.

% SECTION BEYOND HARTREE

\subsection{Numerics at NLO\label{NLO}}
Whereas the LO correlator only knows about the massless (Goldstone)
modes, at NLO the correlator holds information of both the massive
(Higgs) mode and the massless modes. The correlator
turns out to be the average of a massive correlator and $N-1$ massless
correlators (at late times) \cite{Arrizabalaga:2004iw}. In this sense, NLO is a great
improvement on LO. 

We carried out simulations for $N=1,2,4$ and $16$ using both the full classical equations and the NLO 2PI approach. The parameters were identical in both sets of simulations: $\lambda=0.6$, $\mu^2=0.49$, $\Gamma=0.02$, $\delta t=0.1$, and the lattice size was $L=256$ (in lattice units).
In the classical simulations, we averaged over 2000 different initial conditions. This was enough to reduce the statistical error to such a low level that it is impossible to see the error bars in any of our plots.

\FIGURE{
\begin{tabular}{cc}
\epsfig{file=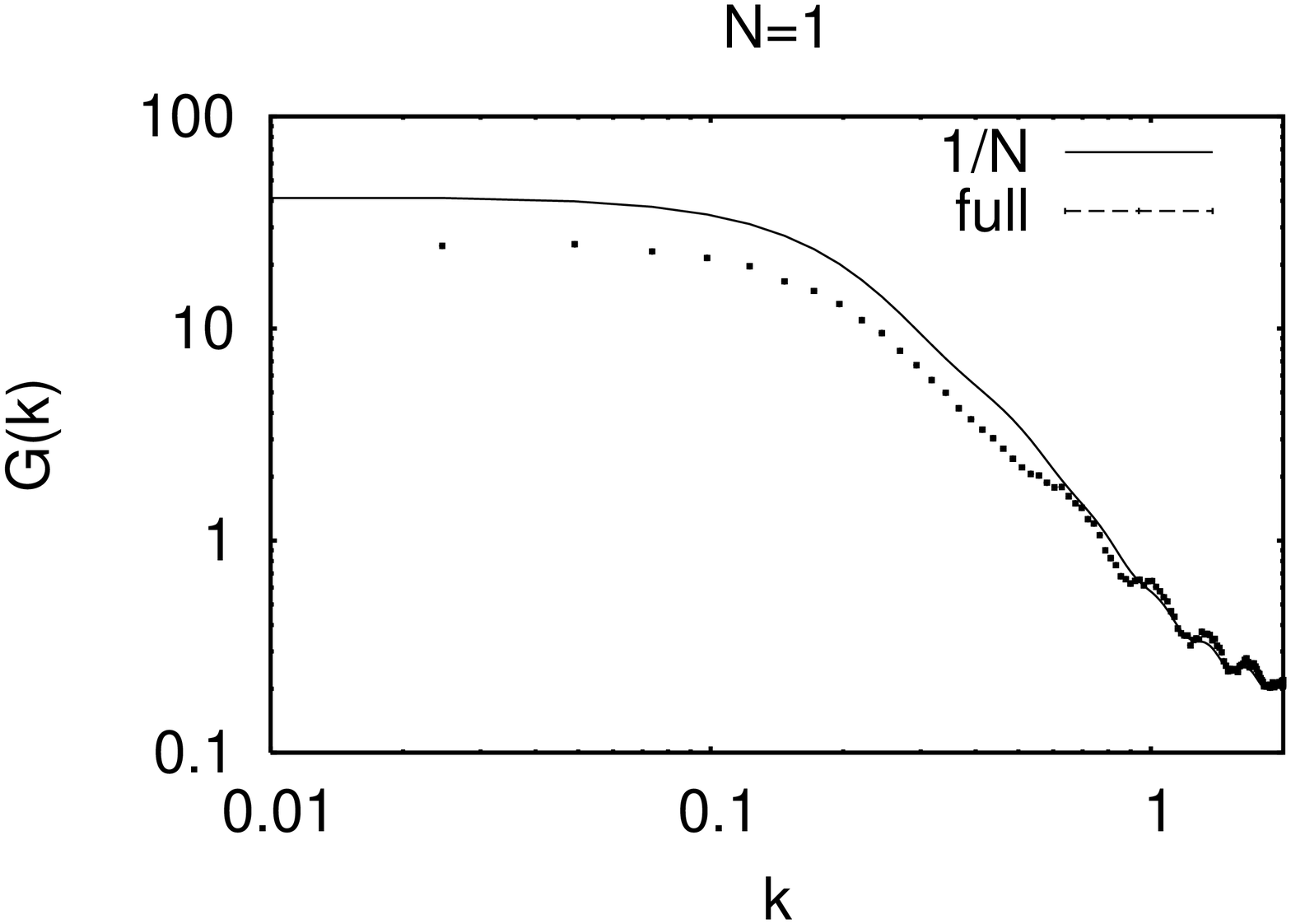,width=7cm,clip}&
\epsfig{file=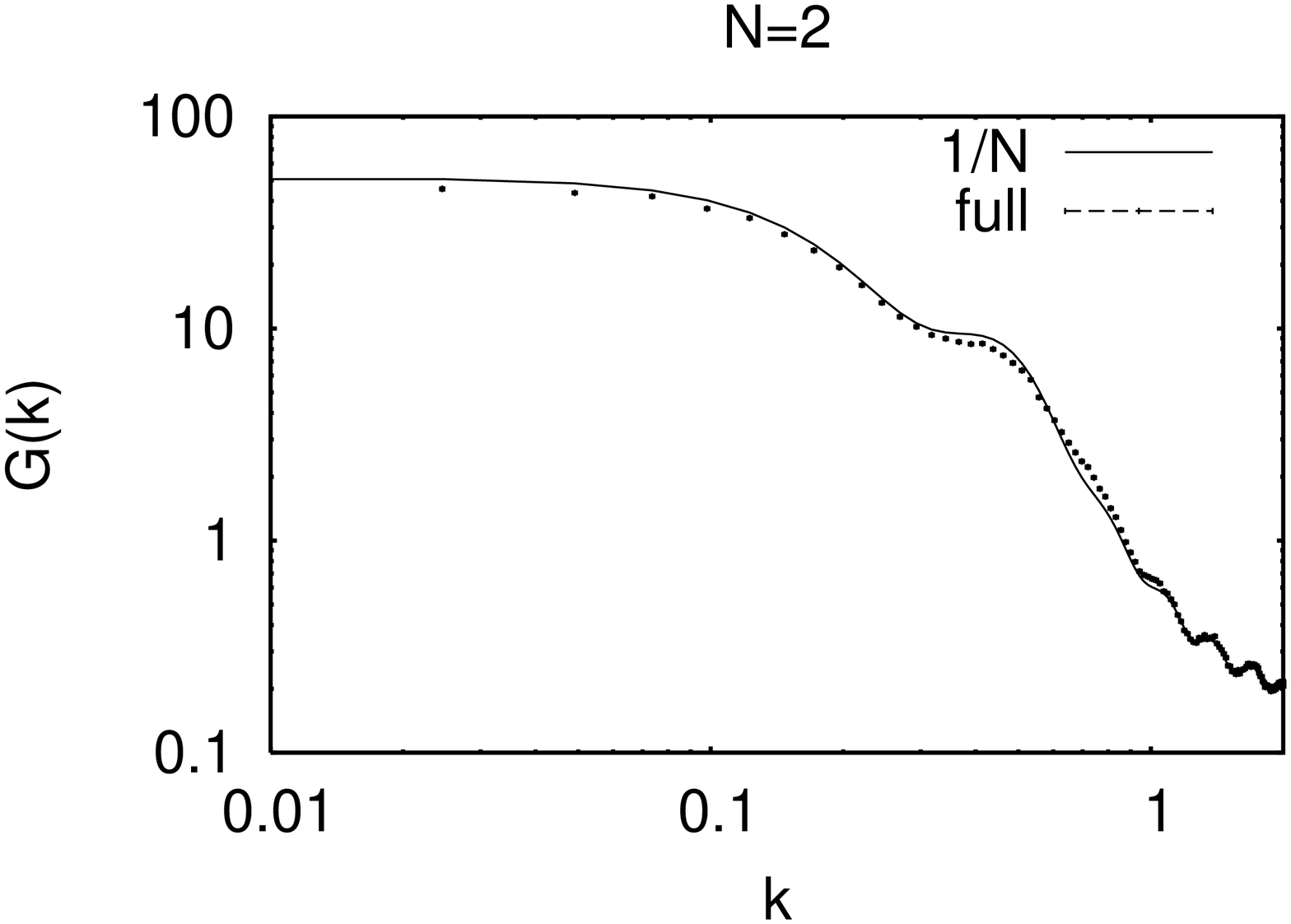,width=7cm,clip}\cr
\epsfig{file=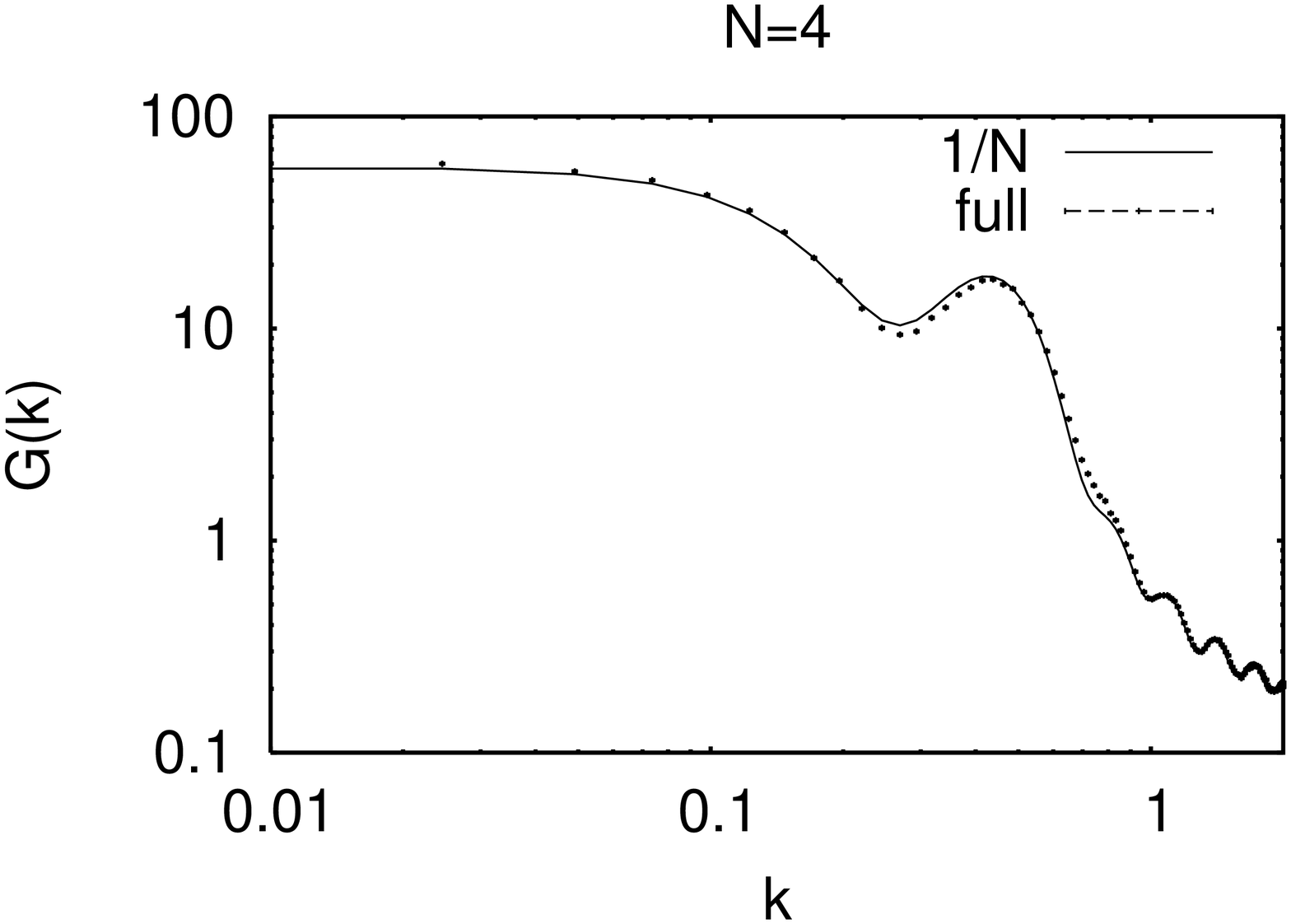,width=7cm,clip}&
\epsfig{file=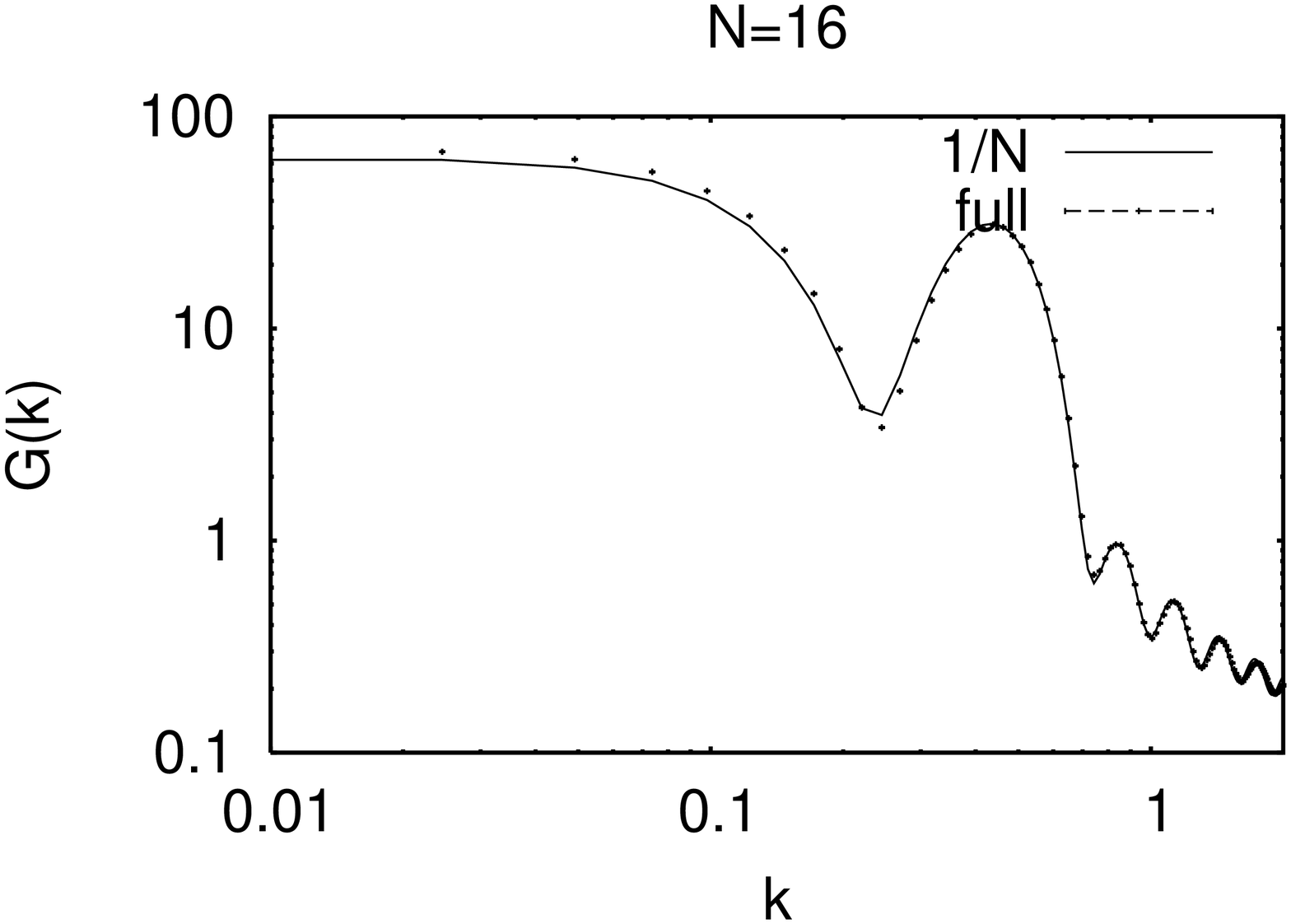,width=7cm,clip}
\end{tabular}
\caption{The equal-time momentum-space correlator $G(k)$ at an early time $t=10$ in the
full classical simulation (dots) and the 2PI-NLO approximation
(line) for different $N$. The parameters were $\lambda=0.6$, $\mu^{2}=0.49$, $\Gamma=0.02$. The agreement is very good for $N>1$ and reasonable even for $N=1$.}
\label{2PI_1}
}

Fig.~\ref{2PI_1} shows the equal-time momentum-space two-point function $G(k)$ at an early time $t=10$. Apart from a small difference at low $k$ in the $N=1$ data, the 2PI approach reproduces the classical results very well. 

\FIGURE{
\begin{tabular}{cc}
\epsfig{file=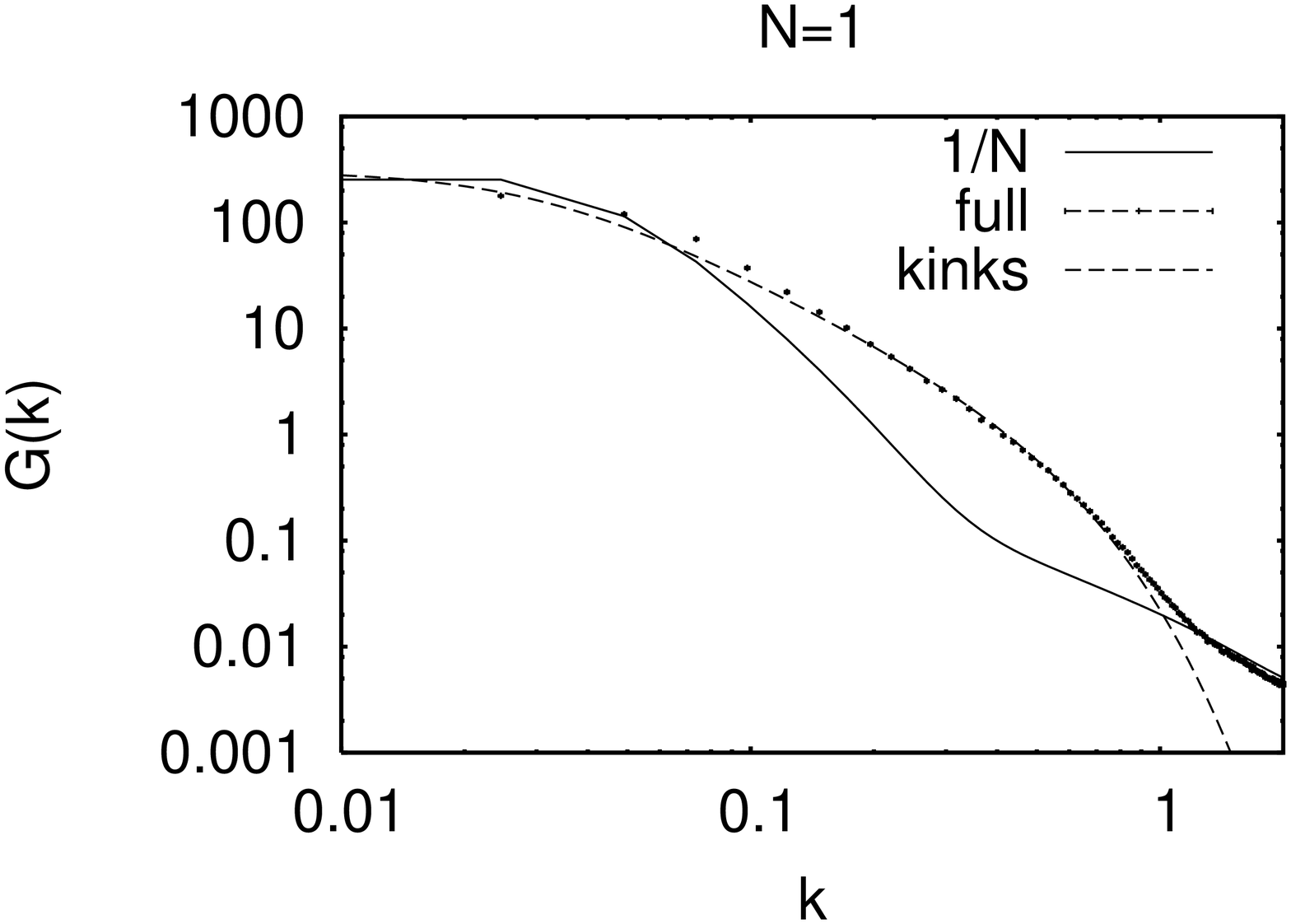,width=7cm,clip}&
\epsfig{file=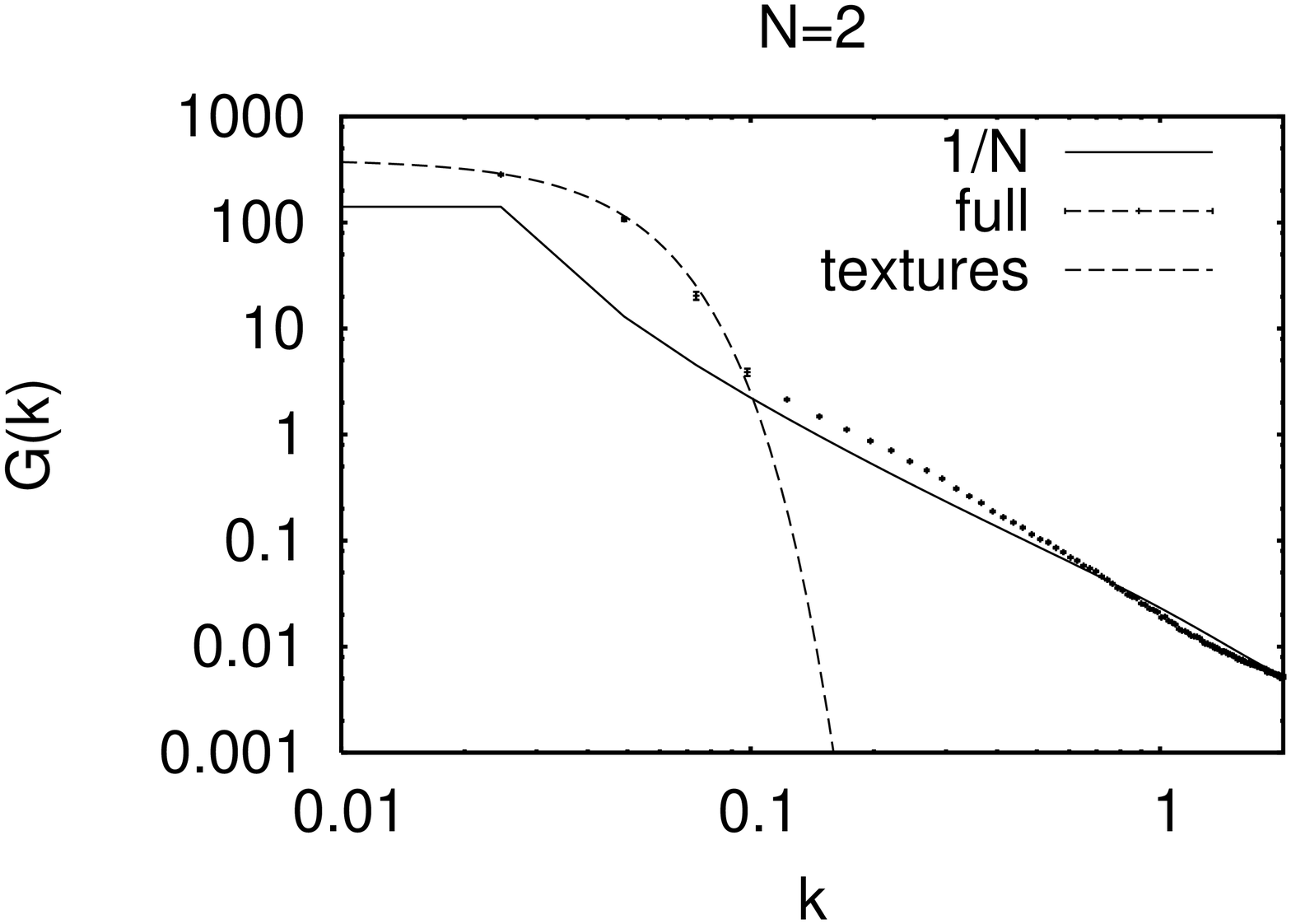,width=7cm,clip}\cr
\epsfig{file=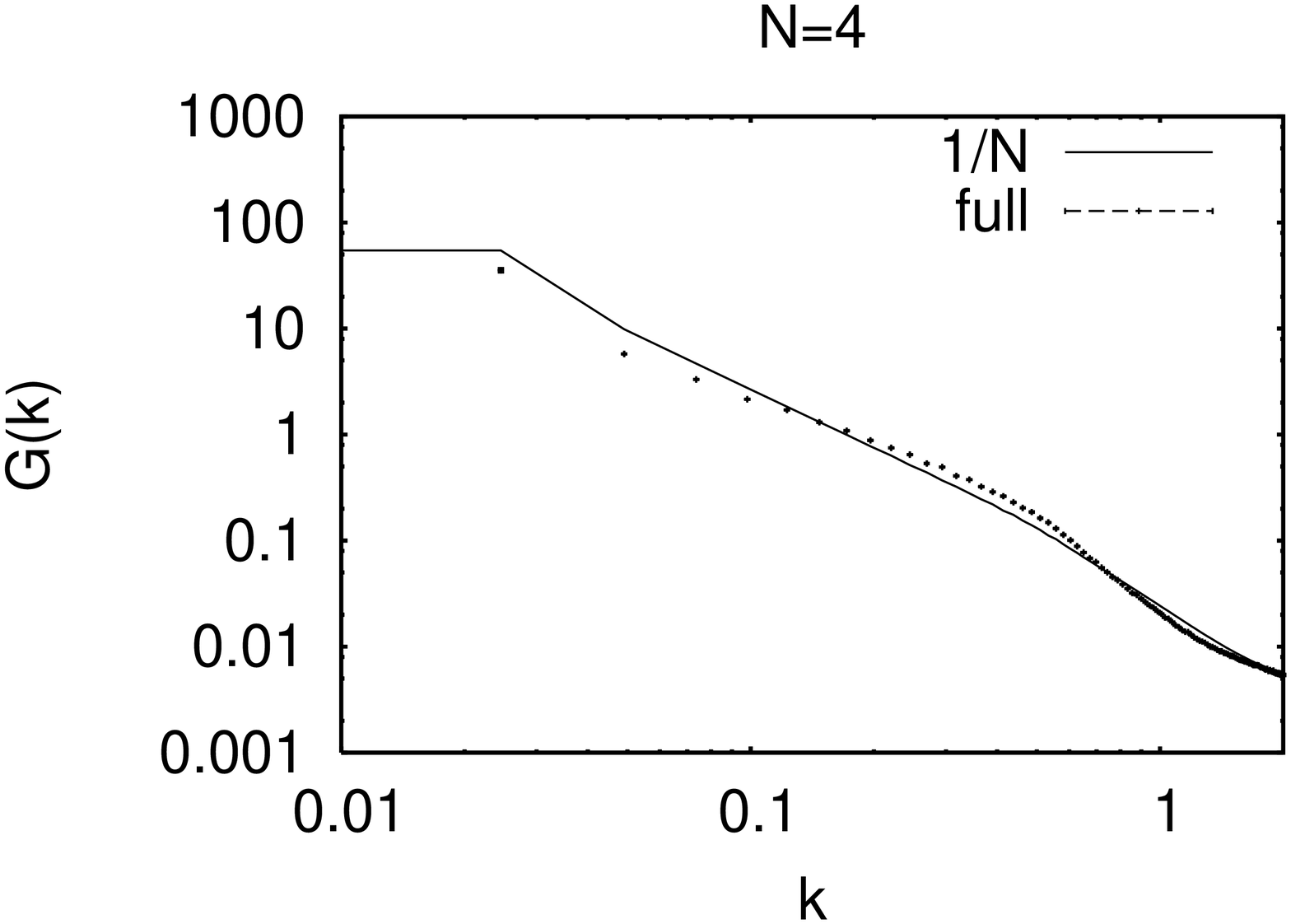,width=7cm,clip}&
\epsfig{file=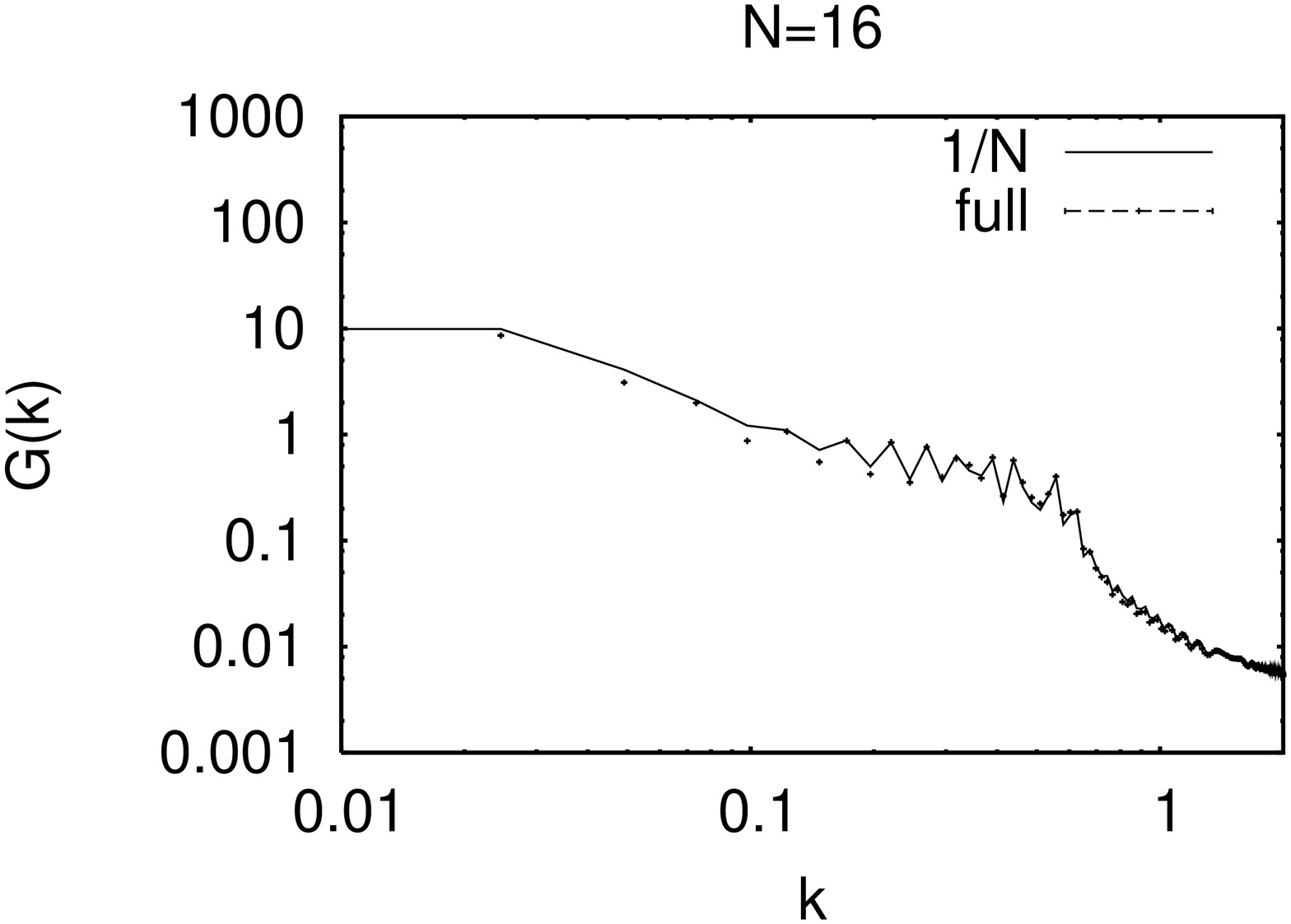,width=7cm,clip}
\end{tabular}
\caption{The equal-time momentum-space correlator $G(k)$ at a late time $t=190$ in the
full classical simulation (dots) and the 2PI-NLO approximation
(line) for different $N$. The parameters were $\lambda=0.6$,
$\mu^{2}=0.49$, $\Gamma=0.02$. The agreement is very good for $N=4$
and $N=16$, but for $N=1$ and $N=2$ there is a clear discrepancy,
which matches precisely  the predicted defect contribution (dashed lines).}
\label{2PI_2}
}

In Fig.~\ref{2PI_2}, we show the corresponding plots at a later time $t=190$, when the system is approaching the equilibrium state. While $N=4$ and $N=16$ still agree very well, there is a clear discrepancy between the
NLO 2PI and the full result for $N=1$ and $N=2$. 

For $N=1$ we did a one-parameter fit to the full data using Eq.~(\ref{equ:kinksign})
with the kink density $n$ as the only free parameter. As the plot shows, this fits the data very well with $n=0.016$. Therefore, we have to conclude that the contribution from the kinks is missing from the NLO 2PI result.

Similarly, we fitted Eq.~(\ref{equ:textsign}) to the $N=2$ data using the length scale $\xi$ as the only free parameter, and obtained a very good fit with $\xi=13.0$. The hump in the correlator at low $k$ is therefore due to textures. Again, there is no sign of this contribution in the NLO 2PI data.

% SECTION OUTLOOK

\section{Conclusions\label{outlook}}

We have shown that the 2PI formalism at next-to-leading order in
a $1/N$ expansions fails to describe topological defects, which exist
in the O($N$) scalar field theory when $N\le 2$. This is
disappointing, because this method has been seen as a promising way of
addressing generic non-equilibrium questions in quantum field
theory. Instead, one has to be very careful when using the method and
make sure that all relevant effects are included, especially since 
there is no hint in the NLO 2PI results themselves that they break down.

One possible caveat is that we are applying a $1/N$ expansion at low
$N$. Still, the fact that the discrepancy is qualitative at both $N=1$
and $N=2$, specifically in the momentum range sensitive to defects, and small already at $N=4$, leads us
to believe that the shortcomings are not a result of the choice
of expansion. Going to higher orders in the $1/N$ approximation
\cite{Aarts:2006cv} is not likely to improve the situation, 
because each type of defects is specific to one particular value of $N$,
and their contribution is therefore non-analytic in $1/N$.
Unfortunately, the obvious alternative, the loop expansion, is in fact
numerically unstable in the presence of large occupation numbers as in the present case of a spinodal transition. 

What this means for the 2PI formalism is that as for any perturbative
framework, certain observables are beyond the reach of the
approximation. One should be careful when studying (symmetry breaking)
phase transitions, and be aware that quantities that are sensitive to the presence of defects will come out wrong. Indeed, since
defects (monopoles, vortices, strings, textures, instantons)
contribute to the propagator even in equilibrium, certain phenomena
may not be reproduced by the 2PI formalism. 
Since the 
expansions on which
 current methods are based  are unlikely to describe topological defects correctly, one may need a radically different approach, perhaps 
along the lines of the Hartree ensemble method developed in Ref.~\cite{Salle:2000hd}.

\subsection*{Acknowledgments}
We thank Gert Aarts, Mark Hindmarsh, Jan Smit and Alejandro Arrizabalaga for
useful discussions. A.T. is supported by PPARC SPG {\it ``Classical lattice field theory''}.

\appendix
\section{Free field\label{appfree}}

It is useful to take a step back and take a look at the simplest
approximation to the dynamics, the free theory. This corresponds to
very early times, as the potential is quenched from positive to
negative curvature. The field experiences a spinodal instability, and
we can solve exactly for the field evolution and the correlator in the potential
\be
\label{freepot}
V(\phi_{a})=V_{0}-\frac{1}{2}\mu^{2}\phi_{a}\phi_{a}.
\ee
The classical and quantum equations coincide, and we can think in
terms of the free Klein-Gordon equation, which reads
\be
\label{kleingordon}
\left(\partial_{t}^{2}+\Gamma\,\partial_{t}+\omega^{-,2}_{k}\right)\phi_{\bf
k}=0,
\ee
with $\omega_{k}^{\pm}=\sqrt{k^{2}\pm\mu^{2}}$. The solution is of the form
\be
\label{phitmore}
\phi_{\bf k}(t)=e^{-\Gamma 
t/2}\left(\alpha_{\bf k}e^{i\,\omega_{k}t}+\beta_{\bf k}e^{-i\,\omega_{k}t}\right),
\ee
with the momentum
\be
\label{pitmore}
\pi_{\bf k}(t)=\partial_{t}\phi_{\bf k}(t)=i\,\omega_{k}e^{-i\Gamma 
t/2}\left[\left(1+i\tilde{\Gamma}\right)\alpha_{\bf
k}e^{i\omega_{k}t}-\left(1-i\tilde{\Gamma}\right)\beta_{\bf
k}e^{-i\omega_{k}t}\right].
\ee
We use
\ba
\label{omegas}
\omega_{k}^{2}=k^{2}-\left(\mu^{2}+\Gamma^{2}/4\right),\qquad
\tilde{\omega}_{k}^{2}=k^{2}+\left(\mu^{2}+\Gamma^{2}/4\right),\qquad \tilde{\Gamma}=\frac{\Gamma}{2\,\omega_{k}}.
\ea
Before the quench at $t<0$, the system is in the vacuum state in the potential $V=V_{0}+\frac{1}{2}\mu^{2}\phi_{a}\phi_{a}$, 
so that
\be
\label{phitless}
\phi_{\bf k}(0^{-})=\frac{1}{\sqrt{2\,\omega_{k}^{+}}}(a_{\bf k}+a_{-\bf
k}^{\dagger}),~~\pi_{\bf
k}(0^{-})=\frac{i\omega_{k}^{+}}{\sqrt{2\,\omega_{k}^{+}}}(a_{\bf k}-a_{-\bf k}^{\dagger}),
\ee
with $\langle a_{\bf k}a_{\bf l}^{\dagger}\rangle=\delta^{3}({\bf k-l})$,
and all other correlators of $a$'s and $a^{\dagger}$'s are zero.

Matching Eqs.~(\ref{phitmore}) and (\ref{pitmore}) to (\ref{phitless}) at $t=0$, it
is then straightforward to calculate the correlators of interest,
\ba
\label{freeprop}
\langle\phi_{\bf k}(t)\phi_{\bf k}^{\dagger}(t')\rangle=
\frac{e^{-\Gamma (t+t')/2}}{4\,\omega_{k}^{+}}&&
\left[
\left(1+\frac{\tilde{\omega}_{k}^{2}}{\omega_{k}^{2}}\right)\cos[\omega_{k}(t-t')]+
\left(1-\frac{\tilde{\omega}_{k}^{2}}{\omega_{k}^{2}}\right)\cos[\omega_{k}(t+t')]
\right.\nonumber\\
&&\left.
-2i\left(\frac{\omega^{+}_{k}}{\omega_{k}}\sin[\omega_{k}(t-t')]+i\tilde{\Gamma}\sin[\omega_{k}(t+t')]\right)
\right],\\
\langle\phi_{\bf k}(t)\pi_{\bf k}^{\dagger}(t')-\pi_{\bf k}(t')\phi_{\bf k}^{\dagger}(t)\rangle=&&
-ie^{-\Gamma (t+t')/2}
\left[\cos[\omega_{k}(t-t')]+\tilde{\Gamma}\sin[\omega_{k}(t-t')]\right]\\
\langle\phi_{\bf k}(t)\pi_{\bf k}^{\dagger}(t')+\pi_{\bf k}(t')\phi_{\bf k}^{\dagger}(t)\rangle=&&
\frac{\omega_{k}\,e^{-\Gamma (t+t')/2}}{2\,\omega_{k}^{+}}
\left[
-\tilde{\Gamma}
\left(1+\frac{\tilde{\omega}_{k}^{2}}{\omega_{k}^{2}}\right)\left(\cos[\omega_{k}(t-t')]-\cos[\omega_{k}(t+t')]\right)
\right.\nonumber\\
&&\left.
+\left(1+\frac{\tilde{\omega}_{k}^{2}}{\omega_{k}^{2}}\right)\sin[\omega_{k}(t-t')]-
\left(1-\frac{\tilde{\omega}_{k}^{2}}{\omega_{k}^{2}}+2\tilde{\Gamma}^{2}\right)\sin[\omega_{k}(t+t')]
\right],\nonumber\\
\ea
\ba
\langle\pi_{\bf k}(t)\pi_{\bf k}^{\dagger}(t')\rangle&&=
\frac{\omega_{k}^{2}\,e^{-\Gamma (t+t')/2}}{4\,\omega_{k}^{+}}
\left[
\left(\tilde{\Gamma}^{2}+1\right)\left(\left(1+\frac{\tilde{\omega}_{k}^{2}}{\omega_{k}^{2}}\right)\cos[\omega_{k}(t-t')]
+\frac{2i\omega_{k}^{+}}{\omega_{k}}\sin[\omega_{k}(t-t')]\right)
\right.\nonumber\\
&&\left.
+
2\tilde{\Gamma}\left(\tilde{\Gamma}^{2}-\frac{\tilde{\omega}^{2}_{k}}{\omega_{k}^{2}}\right)\sin[\omega_{k}(t+t')]+
\left(
\left(
\tilde{\Gamma}^{2}-1
\right)
\left(1-\frac{\tilde{\omega}_{k}^{2}}{\omega_{k}^{2}}\right)
-4\tilde{\Gamma}^{2}
\right)\cos[\omega_{k}(t+t')]
\right],\nonumber\\
\ea
It is easy to convince oneself that these correlators match the results of
\cite{Smit:2002yg} in the limit $\Gamma=0$, $t=t'$. 

Modes with $k>\sqrt{\mu^{2}+\Gamma^{2}/4}$ simply oscillate in
time. For modes with $k<\sqrt{\mu^{2}+\Gamma^{2}/4}$, the oscillation
turns into exponential growth (for large $t\pm t'$),
\be
\label{unstable}
\langle\phi_{\bf k}\phi_{\bf k}^{\dagger}\rangle(t)\propto e^{-\Gamma(t+t')/2} \left(e^{|\omega_{k}|(t+t')}\pm e^{|\omega_{k}|(t-t')}\right).
\ee
For equal time $t=t'$, we have
\be
\label{eqtimeunstable}
\langle\phi_{\bf k}\phi_{\bf
k}^{\dagger}\rangle(t)\propto\exp\left(2\,t\,[\sqrt{\mu^{2}+\Gamma^{2}/4-k^{2}}-\Gamma/2]\right).
\ee
So for all $\Gamma$, modes with $k<\mu$ (rather than
$k<\sqrt{\mu^{2}+\Gamma^{2}/4}$) are unstable. Also, 
\ba
\label{dampeffect}
\sqrt{1+\left(\frac{\Gamma}{2\omega_{k}^{-}}\right)^{2}}-\frac{\Gamma}{2\omega_{k}^{-}}&\rightarrow&
1,\quad \Gamma\rightarrow 0,~~\langle\phi_{\bf k}\phi_{\bf
k}^{\dagger}\rangle(t)\propto e^{2\,|\omega^{-}_{k}|\,t},\\
&\rightarrow&
0,\quad \Gamma\rightarrow \infty,~~\langle\phi_{\bf k}\phi_{\bf
k}^{\dagger}\rangle(t)\propto {\rm constant}.
\ea
It is of course straightforward to generalise the initial state to a state with arbitrary quasi-particle numbers $n_{k}$,
\ba
\label{generalinit}
\langle a_{\bf k}a^{\dagger}_{\bf k}\rangle=n_{k}+1,\quad \langle a^{\dagger}_{\bf k}a_{\bf k}\rangle=n_{k}, 
\ea
for instance starting from a thermal initial state  
\ba
\label{thermalinit}
n_{k}=\left(e^{\omega_{k}^{+}/T}-1\right)^{-1}, ~~{\rm quantum},\qquad
n_{k}=T/\omega_{k}^{+},~~{\rm classical}.
\ea

\section{2PI equations of motion\label{app2PI}}

The 2PI equations result from variation of the 2PI effective action,
which reads \cite{Cornwall:1974vz} (suppressing space labels)
\be
\Gamma[G]=\frac{i}{2}\Tr\ln G^{-1}+\frac{i}{2}G^{-1}_{0}(G-G_{0})+\Gamma_{2},
\ee
where 
\be
G_{0}^{-1}(t,t')=\frac{\delta^{2}(-iS_{0})}{\delta\phi(t)\delta\phi(t')}.
\ee
and $S_{0}$ is the free action. $i\Gamma_{2}$ is the sum of vacuum
diagrams in terms of the Feynman rules, starting at two loops. Extremisation gives the physical propagator
\be
\frac{\delta\Gamma[G]}{\delta G(t,t')}=0\rightarrow
G^{-1}(t,t')=G_{0}^{-1}-2i\frac{\delta\Gamma_{2}}{G(t,t')}.
\ee
Multiplying from the right by $G$, we have
\be
\int_{\mathcal{C}} dt''\left(G_{0}^{-1}(t,t'')-\Sigma(t,t'')\right)G(t'',t')=\delta(t-t'),
\ee
where all integrals are to be performed along the Keldysh contour
$\mathcal{C}$. In the case of our action Eq.~(\ref{action2})
\be
G_{0}^{-1}(t,t')=ia(t)\left(\partial_{t}^{2}-\partial_{x}^{2}+\Gamma\,\partial_{t}+\mu^{2}\right).
\ee
The coupling is effectively time-dependent $\lambda(t)=\lambda a(t)$,
so that the self-energy is given by
\be
\Sigma(t,t')=\frac{\delta\,2i\Gamma_{2}[G,\lambda(t)]}{\delta G(t,t')},
\ee
For the purpose of illustration, we will look at the coupling
expansion. At order $\lambda$, the effective action contains a single
diagram, the ``figure-8'',
\be
i\Gamma^{8}_{2}=-i\lambda(t)\int_{\mathcal{C}} dt\,G_{ab}^{2}(t,t).
\ee
This leads to the Hartree approximation, for which we find
\be
a(t)\left(\partial_{t}^{2}-\partial_{x}^{2}+\Gamma\partial_{t}-\mu^{2}\right)G(t,t')=-\lambda(t)\frac{N+2}{6N}F(t,t)G(t,t').
\label{hartreeeom}
\ee
In this case, $a(t)$ cancels out, and the equation of motion
reduces to the usual Hartree approximation, with an added
damping term. This makes perfect sense, since at Hartree order,
classical and quantum evolution coincides, and $a(t)$ appears only in
the classical equation of motion Eq.~(\ref{equ:eom}) through the damping term.

Beyond the Hartree order, the modification by including the expansion
is to make the replacement $\lambda\rightarrow\lambda a(t)$ for all
vertices of the self-energy. So, for the sunset diagram (order
$\lambda^{2}(t)$) we should add
\ba
\label{sunset1}
i\Gamma^{\rm sunset}_{2}= \frac{(-i\lambda)^{2}}{48}\int_{\mathcal{C}}dt\,dt'\,a(t)G^{4}(t,t')a(t'),\\  
\Sigma^{\rm sunset}(t,t')=-\frac{\lambda^{2}}{6} a(t)G(t,t')^{3}a(t').
\ea
In the equation of motion, we add to the right hand side of Eq.~(\ref{hartreeeom})
\be
\label{sunset2}
\int dt'' \Sigma(t,t'')G(t'',t')= -\frac{\lambda^{2}a(t)}{6}\int dt''\,G(t,t'')^{3}a(t'')G(t'',t').
\ee
Note, that a factor of $a(t)$ will again cancel out, leaving only the
$a(t'')$ which is integrated over.

An alternative interpretation of this is to return to the original 2+1
dimensional system Eq.~(\ref{action}). If we consider the scale
factor $a(t)$ as part of the metric rather than the Lagrangian, then
$\lambda$ is time-independent, and the integrals become
\be
\label{altinterp}
\int dt''\,dx\,a(t'')dy \lambda^{2}f(t,t',t'',x,y)= \int dt''\,dx L_{y}a(t) \lambda^{2}f(t,t',t'',x),
\ee
so that the occurrence of $a(t'')$ is a result of integrating over a
homogeneous slice in $y$, with increasing size $L_{y}a(t'')$.

We can attempt a naive $a(t)$-power counting in perturbation
theory. The free propagator Eq.~(\ref{freeprop}) is of the form
\be
\label{freeorder}
G(t,t',a(t),a(t'))= \frac{\tilde{G}(t,t')}{\sqrt{a(t)a(t')}}\propto \frac{1}{a(t)}\sqrt{\frac{a(t)}{a(t')}},
\ee
neglecting the instability of some of the modes, for which the growth is
determined by $k^{2}-\mu^{2}$ and not $a(t)$.

We can now write the Schwinger-Dyson equation in powers of $a(t)$ by
assuming that the full propagator supplies the same power of
$a(t)$ as the free $G(t,t')\propto a^{-1}(t)(a(t)/a(t'))^{1/2}$, 
\ba
\label{powercount}
\left(O[{\bf Free}]+O[{\bf Hartree}]+O[{\bf
NLO}]\right)G(t,t')=\delta(t-t'),\nonumber\\
\\
\left(O[a(t)^{1}]+O[a^{0}(t)]\right)G(t,t')+\int
dt''O[a^{-1}(t)\sqrt{\frac{a(t)}{a(t'')}}]G(t'',t')=\delta(t-t').\nonumber\\
\ea
For each extra vertex, we get another order $a^{-1}$. This is because
each extra vertex supplies one factor of $\lambda(t''')$ and four extra
propagators legs (for a four-vertex), giving
$(a^{-1/2}(t'''))^{4}a(t''')$. In this way, each subsequent order of
$\lambda$ gives an extra integration and a factor of $a^{-1}$. In
contrast, had we simply counted orders of $\lambda(t)$ without
reference to the suppression of the propagators, we would have had a
factor of $a(t)$ for each extra vertex, which in this case is
exponentially growing in time, making the expansion unreliable at best.

In practice, it turns out that the $1/N$ expansion is more robust
towards very large occupation numbers, as in the present case of a
spinodal instability. The inclusion of the expansion in the $1/N$ case
proceeds along the same lines as for the coupling expansion discussed
here, resulting in Eqs.~(\ref{schwingerdyson}-\ref{sigmaNLO}).

\bibliography{andersbib}

\end{document}

%% file: newcomm.tex
\newcommand{\Tr}{\mbox{Tr}\,}

\newcommand{\eela}[1]{\label{#1}\end{equation}}
\newcommand{\eeala}[1]{\label{#1}\end{eqnarray}}
\newcommand{\bea}{\begin{eqnarray}}
\newcommand{\eea}{\end{eqnarray}}